# Impact of ferroelectric nonlinearity and correlation effects on nanodomain formation

Anna N. Morozovska*

Institute of Semiconductor Physics, 41, pr. Nauki, 03028 Kiev,

National Academy of Science of Ukraine, Ukraine

## Abstract

Using direct variational method with 2-parametric trial function and Landau-Ginzburg-Devonshire thermodynamical approach, we derived analytical expressions for polarization spatial redistribution in the ferroelectrics caused by the biased Scanning Probe Microscope probe. We demonstrate that the shape of nanodomain induced by the probe electric field can be either oblate or prolate depending on the ferroelectric nonlinearity strength. For typical ferroelectric material parameters and probe apex geometry the domain nucleus aspect ratio is close to dielectric anisotropy factor. Corresponding coercive biases of a stable domain formation are in reasonable agreement with available experimental results. Spike-like domains typical for Landauer-Molotskii rigid approach appear in the considered case only when depolarization field energy contribution strongly dominates over the nonlinear correlation and field effects and domain wall energy.

## 1. Overview

The Scanning Probe Microscopy (SPM) based techniques open the way to concentrate electric field within a nanoscale volume of material [1, 2]. Combined with electromechanical response detection, this Piezoresponse Force Microscopy approach has been broadly applied for domain imaging and polarization patterning in ferroelectrics [3, 4]. Piezoresponse force spectroscopy was used to study polarization switching in the small volumes with negligible defect concentration [5], map distribution of random bond- and random field components of disorder potential [6, 7], and map polarization switching on a single defect center [8]. These experimental developments have been complemented by the extensive theoretical analysis of domain nucleation mechanisms in the SPM field probe on the ideal surface [9, 10, 11, 12] and in the presence of charged defects [13] in the rigid approximation.

* Corresponding author: morozo@i.com.ua

The rigid ferroelectric approximation was proposed by Landauer [14] for the domain nucleation in homogeneous electric field of plain capacitor. Then the approach was adopted for the domain formation caused by inhomogeneous electric field of the force microscope charged probe by Molotkii et al [15, 9, 11]. Within rigid Landauer-Molotskii (LM) model domain walls between the regions with field-independent “rigid” spontaneous polarization $\pm P_S$ are regarded atomically-sharp (mathematically infinitely thin).

Landau-Ginzburg-Devonshire (LGD) thermodynamic theory considers continuous spatial distribution of polarization vector $P_i(x,y,z)$ in arbitrary electric field and nonlinear long-range polarization interactions (correlation effects) [16]. So, LGD-approach avoids typical limitations (sharp walls and field-independent polarization value) of rigid approximation.

Intrinsic domain wall width is a fundamental parameter that reflects bulk ferroelectric properties and governs the performance of ferroelectric memory devices [17]. Recently [18] we derived closed-form analytical expressions for vertical and lateral Piezoelectric Force Microscopy (PFM) profiles of a single ferroelectric domain wall taking into account the finite intrinsic width of the domain wall. Note, that depolarization field drastically decreases with the domain wall width increase. Namely, Gopalan et al [19] recently have shown that the anti-parallel counter domain wall width increase from 0.5 nm to 2…3 nm leads to the coercive field decrease on 2-3 orders of magnitude.

In Ref.[20] we consider the interaction of ferroelectric $180^{o}$-domain wall polarization with a strongly inhomogeneous electric field of biased force microscope probe within the LGD-approach for the *second order* ferroelectrics. The approximate analytical expressions for the equilibrium *surface* polarization distribution were derived using direct variational method with *one-parametric* trial function. However, at least ***two-parametric trial functions*** are necessary for more rigorous analytical calculations of polarization profiles [21].

In Ref.[22] thermodynamics of tip-induced nanodomain formation in scanning probe microscopy of ferroelectric films and crystals was studied using the analytical Landau-Ginzburg-Devonshire *local* approach and phase-field modeling. The *local redistribution* of polarization induced by the biased probe apex is analyzed including the effects of polarization gradients, field dependence of dielectric properties, intrinsic domain wall width, and film thickness. The polarization distribution inside a "subcritical" nucleus of the domain preceding the nucleation event is shown to be "soft" (i.e. smooth without domain walls) and localized below the probe. Polarization distribution inside a stable domain is "hard" and the spontaneous polarization reorientation takes place inside a localized spatial region, where the absolute value of the resulting electric field is larger than the thermodynamic coercive field.

In the present study we combined LGD theory with direct variational method for description of the stable nanodomain formation. using direct variational method with two-parametric trial function we derived analytical expressions appropriate for both *first* and *second order* ferroelectrics. The expressions provide insight how the polarization re-distribution depends on the wall finite-width, nonlinear correlation and depolarization effects, electrostatic potential distribution of the probe and ferroelectric material parameters.

## 2. The problem statement

The spontaneous polarization $P_3(\mathbf{r})$ of ferroelectric is directed along the polar axis, *z*. The sample is dielectrically isotropic in transverse directions, i.e. permittivities $\varepsilon_{11}=\varepsilon_{22}$, while $\varepsilon_{33}$ value may be different. The dependence of in-plane polarization components on electric field is linearized as $P_{1,2}\approx-\varepsilon_0(\varepsilon_{11}-1)\partial\varphi(\mathbf{r})/\partial x_{1,2}$. Then the problem for the electrostatic potential $\varphi(\mathbf{r})$ acquires the form [22]:

$$\begin{cases}\varepsilon_{33}^{b}\dfrac{\partial^2\varphi}{\partial z^2}+\varepsilon_{11}\left(\dfrac{\partial^2\varphi}{\partial x^2}+\dfrac{\partial^2\varphi}{\partial y^2}\right)=\dfrac{1}{\varepsilon_0}\dfrac{\partial P_3}{\partial z},\\ \varphi(x,y,z=0)=V_e(x,y),\quad \varphi(x,y,z=h)=0\,.\end{cases}\tag{1}$$

Here we introduced dielectric permittivity of background [23] as $\varepsilon_{33}^{b}$. Typically $\varepsilon_{33}^{b}\leq10$. $V_e(x,y)$ is the potential distribution at the sample surface; $\varepsilon_0$ is the universal dielectric constant, *h* is the sample thickness.

Electrostatic potential $\varphi(\mathbf{r})$ includes the effects of depolarization field created by polarization bound charges. The perfect screening of depolarization field outside the sample is realized by the ambient screening charges.

Corresponding Fourier-image on transverse coordinates $\{x,y\}$ of electric field normal component $\widetilde{E}_3(\mathbf{k},z) = -\partial\widetilde{\varphi}/\partial z$ is the sum of external (*e*) and depolarization (*d*) fields [20, 22]:

$$\widetilde{E}_3(\mathbf{k},z) = \widetilde{E}_3^e(\mathbf{k},z) + \widetilde{E}_3^d(\mathbf{k},z), \tag{2a}$$

$$\widetilde{E}_3^e(\mathbf{k},z) = \widetilde{V}_e(\mathbf{k})\frac{\cosh(k(h-z)/\gamma_b)}{\sinh(k h/\gamma_b)}\frac{k}{\gamma_b}, \tag{2b}$$

$$\widetilde{E}_3^d(\mathbf{k},z) = \left(\begin{array}{l} \int_0^z dz'\frac{\widetilde{P}_3(\mathbf{k},z')}{\varepsilon_0\varepsilon_{33}^b}\cosh(k z'/\gamma_b)\frac{\cosh(k(h-z)/\gamma_b)}{\sinh(k h/\gamma_b)}\frac{k}{\gamma_b} + \\ \int_z^h dz'\frac{\widetilde{P}_3(\mathbf{k},z')}{\varepsilon_0\varepsilon_{33}^b}\cosh(k(h-z')/\gamma_b)\frac{\cosh(k z/\gamma_b)}{\sinh(k h/\gamma_b)}\frac{k}{\gamma_b} - \frac{\widetilde{P}_3(\mathbf{k},z)}{\varepsilon_0\varepsilon_{33}^b} \end{array}\right). \tag{2c}$$

Here $\gamma_b = \sqrt{\varepsilon_{33}^b/\varepsilon_{11}}$ is the "bare" dielectric anisotropy factor, $\mathbf{k} = \{k_1,k_2\}$ is a spatial wave-vector, its absolute value $k = \sqrt{k_1^2 + k_2^2}$. For a transversally homogeneous media, $\varepsilon_{33}^b = 1$ and static case Eq. (2c) reduces to the expression for depolarization field obtained by Kretschmer and Binder [24].

Potential distribution produced by the SPM probe on the surface of semi-infinite sample can be approximated as $V_e(x,y) \approx V d/\sqrt{x^2 + y^2 + d^2}$, where $V$ is the applied bias, $d$ is the effective distance determined by the probe geometry (see Ref.[12] and Fig. 1). Corresponding Fourier-image on transverse coordinates $\{x,y\}$ of electric field potential at the sample surface is $\widetilde{V}_e(\mathbf{k}) = V\widetilde{w}(\mathbf{k})$, where $\widetilde{w}(\mathbf{k}) = d\exp(-kd)/k$. The potential is normalized assuming the condition of perfect electrical contact with the surface, $V_e(0,0) \approx V$. In the case of local point charge model, the probe is represented by a single charge $Q = 2\pi\varepsilon_0\varepsilon_e R_0 V(\kappa + \varepsilon_e)/\kappa$ located at distance $d = \varepsilon_e R_0/\kappa$ for a spherical tip apex with curvature $R_0$ ($\kappa$ is the effective dielectric constant determined by the "full" dielectric permittivity in z-direction, $\varepsilon_e$ is ambient dielectric constant), or $d = 2R_0/\pi$ for a flattened tip represented by a disk of radius $R_0$ in contact with the sample surface [12].

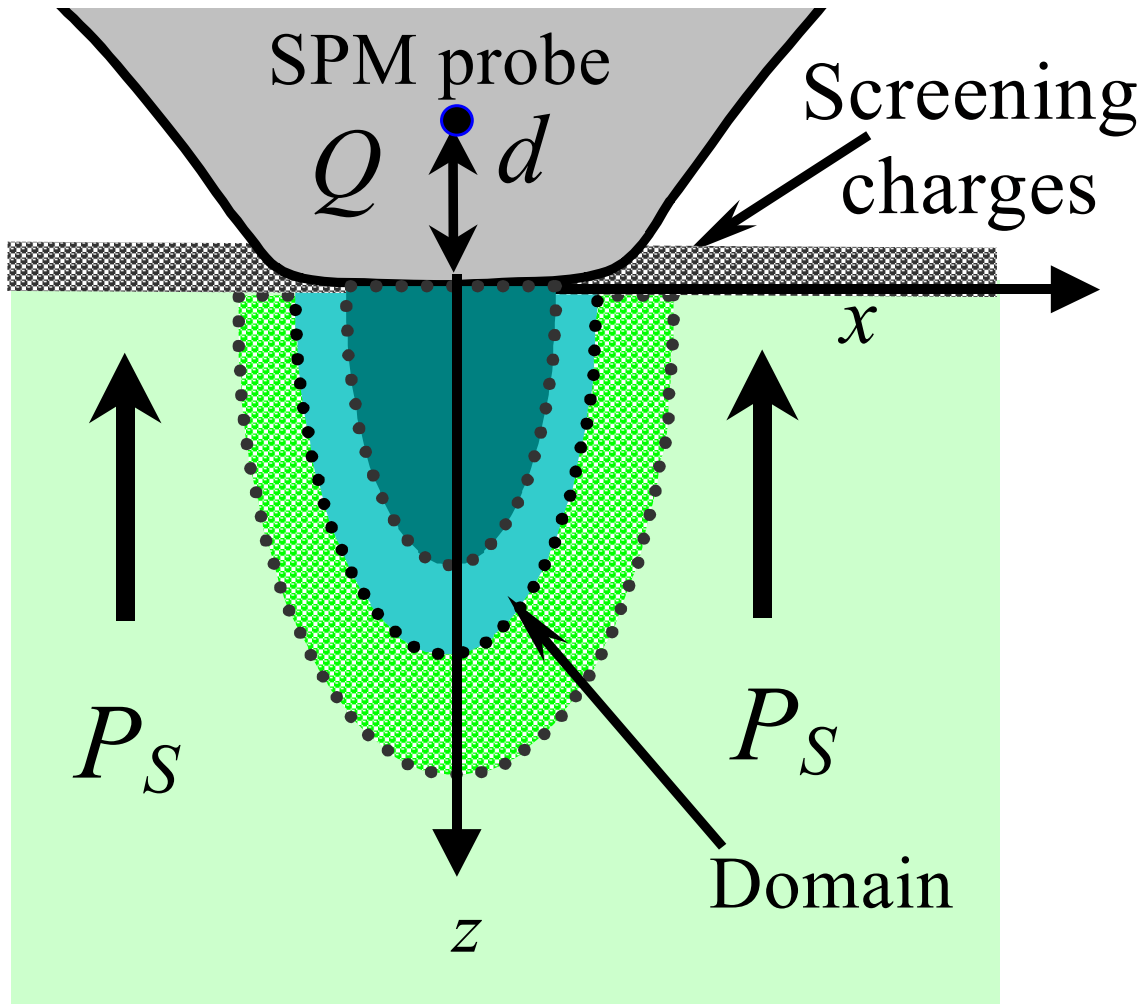

**FIG. 1**. Nanodomain caused by the electric field of the biased SPM probe in contact with the sample surface.

In the framework of LGD phenomenology, the stable or metastable polarization distribution inside the proper ferroelectric can be found as the solution of the nonlinear equation:

$$\alpha P_3 + \beta P_3^3 + \delta P_3^5 - \xi \frac{\partial^2 P_3}{\partial z^2} - \eta \left( \frac{\partial^2 P_3}{\partial x^2} + \frac{\partial^2 P_3}{\partial y^2} \right) = E_3 \ . \tag{3a}$$

The solution is the extremum of the free energy

$$G = \int_{-\infty}^{\infty} dx \int_{-\infty}^{\infty} dy \left( \int_0^h dz \left( \frac{\alpha}{2} P_3^2 + \frac{\beta}{4} P_3^4 + \frac{\delta}{6} P_3^6 + \frac{\xi}{2} \left( \frac{\partial P_3}{\partial z} \right)^2 + \frac{\eta}{2} (\nabla_\perp P_3)^2 - P_3 \left( E_3^e + \frac{E_3^d}{2} \right) \right) \right) \tag{3b}$$

The gradient (or correlation) terms $\xi > 0$ and $\eta > 0$ (usually $\xi \sim \eta$), expansion coefficients $\delta > 0$, while $\beta < 0$ for the first order phase transitions or $\beta > 0$ for the second order ones. Coefficient $\alpha<0$ in ferroelectric phase. Rigorously, coefficient $\alpha$ should be taken as renormalized by the elastic stress [25, 26].

For the semi-infinite sample considered hereinafter, boundary conditions for polarization are the following

$$P_3(r >> d) \to -P_S, \quad \frac{\partial P_3}{\partial z}(z = 0) = 0 \, . \tag{4}$$

$P_S$ is the initial spontaneous polarization value. The condition $\partial P_3/\partial z = 0$ corresponds to the perfect atomic surface structure without defects or damaged layer (for the case one could neglect the surface energy contribution). Constant polarization value $P_S$ satisfies Eq.(3) at zero external bias, $V$=0. For the first order ferroelectric the spontaneous polarization $P_S^2 = \left(\sqrt{\beta^2 - 4\alpha\delta} - \beta\right)/2\delta$, while $P_S^2 = -\alpha/\beta$ for the second order one [16].

## 3. Polarization reversal caused by the biased probe

To obtain the spatial distribution of polarization *near the biased probe apex* we used ***direct variational method with 2-parametric trial function***. Firstly Eq.(3a) was linearized as $P_3(\mathbf{r}) = -P_S + p(\mathbf{r})$, where $p(\mathbf{r})$ is the deviation, we are looking for within perturbation approach [20]. Naturally, the condition $p(\mathbf{r}) \to 0$ should be valid far from the probe at arbitrary applied bias $V$. Then *coordinate-dependent part* of linearized solutions was used as the trial functions in the free energy functional (3b).

Two-parametric trial function is introduced as

$$P_3(\rho, z) = -P_S + p_V \left( \begin{array}{c} \dfrac{(d + z/\gamma_V)d^2}{\left(L_\perp (d + z/\gamma_V) + (d + z/\gamma_V)^2 + \rho^2\right)^{3/2}} \\ + \dfrac{d^2(d^2 + \rho^2) - 3d^4}{\gamma_V (d^2 + \rho^2)^{5/2}} L_z \exp\left(-\dfrac{z}{L_z}\right) \end{array} \right). \qquad (5)$$

Here $\rho = \sqrt{x^2 + y^2}$ has the meaning of radial coordinate. The length $L_\perp = \sqrt{\eta/(\alpha + 3\beta P_S^2 + 5\delta P_S^4)}$ originated from the intrinsic width of domain wall, the correlation length $L_z = \sqrt{\varepsilon_0 \varepsilon_{33}^b \xi}$ is extremely small due to the depolarization effects. When deriving expression (5) for the trial function, we used that the inequalities $2\varepsilon_0 \varepsilon_{33}^b |\alpha| << 1$, $\varepsilon_{33}^b << \varepsilon_{33}$, $L_\perp < 1$ nm and $L_z < 1$ Å are typically valid for ferroelectric material parameters and background permittivity $\varepsilon_{33}^b \le 10$. Hereinafter we use that the inequality $L_z << L_\perp << d$ is valid. It leads to the approximation $P_3(\rho, z) \approx -P_S + p_V \dfrac{(d + z/\gamma_V)d^2}{\left((d + z/\gamma_V)^2 + \rho^2\right)^{3/2}}$.

Effective dielectric anisotropy factor $\gamma_V(V)$ that determines *z*-scale and the amplitude $p_V(V)$ are *variational parameters*, which should be found *self-consistently* from the minimum of the free energy (3b) allowing for the conditions $\gamma_V(0)=\sqrt{\frac{1+\varepsilon_{33}^b\varepsilon_0\left(\alpha+3\beta P_S^2+5\delta P_S^4\right)}{\varepsilon_{11}\varepsilon_0\left(\alpha+3\beta P_S^2+5\delta P_S^4\right)}}\approx\sqrt{\frac{\varepsilon_{33}}{\varepsilon_{11}}}$ and $p_V(0)=0$, since $p_V\sim V$ for small bias.

Substituting the trial function (5) into the free energy functional (3b), after approximate integration we obtained the free energy excess with renormalized coefficients

$$
\begin{gathered}
G(p_V,\gamma_V,V)=G_V(p_V,\gamma_V,V)+G_W(p_V,\gamma_V)+G_D(p_V,\gamma_V)+G_N(p_V,\gamma_V),\\
G_V(p_V,\gamma_V,V)\approx-\frac{Vd^2\gamma_V p_V}{2(\gamma_V+\gamma_b)},\quad G_W(p_V,\gamma_V)\approx\frac{\gamma_V d}{16}\left(\eta+\frac{\xi}{\gamma_V^2}\right){p_V}^2,\\
G_D(p_V,\gamma_V)\approx\frac{\gamma_V d^3{p_V}^2}{8\varepsilon_0\varepsilon_{11}(\gamma_V+\gamma_b)^2},\\
G_N(p_V,\gamma_V)\approx\gamma_V d^3\left(\begin{array}{l}\left(\alpha+3\beta P_S^2+5\delta P_S^4\right)\frac{{p_V}^2}{8}-\left(\beta P_S+\frac{10\delta}{3}P_S^3\right)\frac{{p_V}^3}{21}+\\ \left(\frac{\beta}{50}+\frac{\delta}{5}P_S^2\right)\frac{{p_V}^4}{4}-\delta P_S\frac{{p_V}^5}{91}+\frac{\delta}{144}\frac{{p_V}^6}{6}\end{array}\right).
\end{gathered}
\tag{6}
$$

Here $G_V$ is the interaction energy with probe external field, $G_W$ is the domain wall (or gradient) energy, $G_D$ is depolarization field energy and $G_N$ is nonlinear correlation energy reflecting the field dependence of polarization value. More rigorous, but cumbersome expressions for the free energy of the second order ferroelectrics are listed in the Supplement. Equations of state are $\partial G/\partial\gamma_V=0$, $\partial G/\partial p_V=0$.

Within rigid model domain walls are regarded infinitely thin and polarization absolute *value* is constant: it is $-P_S$ outside and $+P_S$ inside the semi-ellipsoidal domain (if any). Semi-ellipsoidal domain radius *r* and length *l* are calculated from the free energy excess consisting of the interaction energy, the domain wall surface energy and depolarization field energy listed in the Supplement (see Refs. [9], [12], [15]). Let us underline, that nonlinear correlation energy $G_N$ contribution is absent within rigid approximation.

Variational parameters $\gamma_V$ and $p_V$ bias dependence are shown in Figs.2 for typical ferroelectric materials.

It is seen from the Figs.2, that factor $\gamma_V$ *weakly* and *non-monotonically* depends on applied bias $V$ within LGD approach with high values of nonlinear energy $G_N$ in contrast to the rapid increase of $\gamma_V$ calculated for the case $G_N << G_W$ and $p_V(V) \approx 2P_S$. For high $G_N$ the amplitude $p_V$ increases with bias increase in accordance with nonlinear equations of state obtained by the free energy (6) minimization (compare plots (a) and (b)).

Since the domain length $l(V) \sim \gamma_V d$, rapid increase of $\gamma_V$ means spike-like domain formation. It is worth to underline, that spike-like domains are possible only when depolarization field energy contribution strongly dominates over the nonlinear correlation effects and domain wall energy. Within rigid model depolarization field energy vanishes as $1/l$, while the interaction energy is maximal at $l\rightarrow\infty$, the condition of negligible surface energy leads to the domain breakdown $l\rightarrow\infty$ and subsequent macroscopic region re-polarization even at infinitely small bias (if only $VP_S>0$), while hysteresis phenomena or threshold bias (saddle point) are absent. Domain breakdown was calculated by Molotskii et al [9]. The result reflects the conventional statement that under the absence of pinning, *spontaneous* polarization of defect-free ferroelectric can be reversed by infinitely small field applied long enough [27]. In other words, hysteresis phenomenon appearance has to correspond to the *metastable* (*non-ergodic*) *state*.

Note that high PFM response contrast is possible under the condition $|p_V| > 2P_S$ obtained within LGD-approach with applied bias increase. This opens pathway for high-density data storage in ultra-thin layer even for small domain length. Surely, for the case $p_V(V)/P_S >> 1$ higher expansion terms should be considered and polarization value should increase much more slowly.

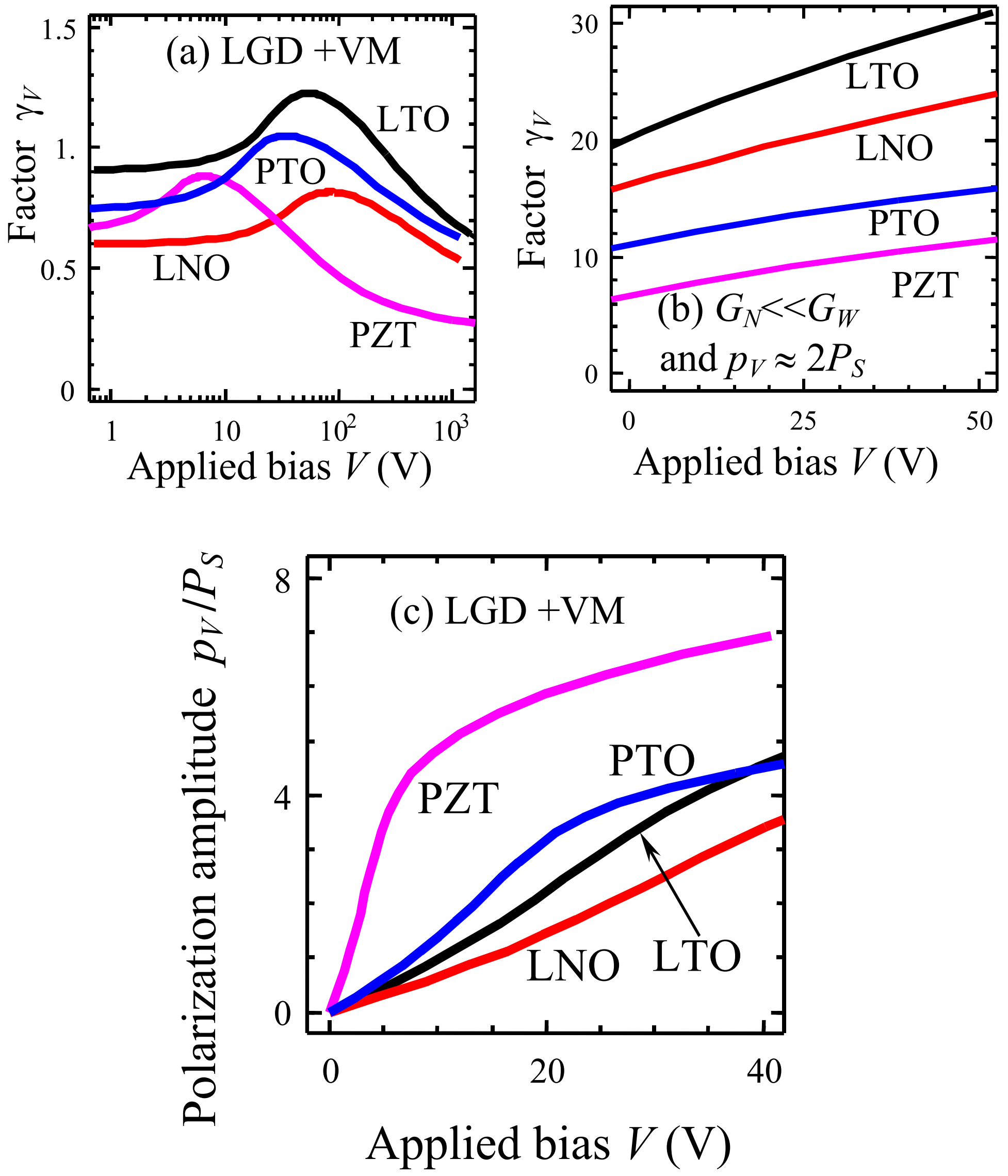

**FIG. 2**. Variational parameters $\gamma_V(V)$ (a,b) and $p_V(V)$ (c) bias dependence calculated with LGD approach combined with 2 parametric direct variational method (LGD+VM) for typical ferroelectric materials: PTO ($PbTiO_3$ with $\varepsilon_{11}$=124, $\varepsilon_{33}$=67, $\alpha = -3.42\cdot10^8$ m/F, $\beta$= -2.90·$10^8$ $m^5/(C^2F)$, $\delta$= 1.56·$10^9$ $m^5/(C^2F)$, $P_S$=0.75 $C/m^2$); PZT ($PbZr_{40}Ti_{60}O_3$ with $\varepsilon_{11}$=497, $\varepsilon_{33}$=197, $\alpha = -1.66\cdot10^8$ m/F, $\beta$= 1.44·$10^8$ $m^5/(C^2F)$, $\delta$= 1.14·$10^9$ $m^5/(C^2F)$, $P_S$=0.57 $C/m^2$) [28]; LTO ($LiTaO_3$ with $\varepsilon_{11}$=54, $\varepsilon_{33}$=44, $\alpha = -1.31\cdot10^9$ m/F, $\beta$= 5.04·$10^9$ $m^5/(C^2F)$, $P_S$=0.51 $C/m^2$); LNO ($LiNbO_3$ with $\varepsilon_{11}$=84, $\varepsilon_{33}$=30, $\alpha = -1.95\cdot10^9$ m/F, $\beta$= 3.61·$10^9$ $m^5/(C^2F)$, $P_S$=0.73 $C/m^2$) [29]. Gradient coefficients $\eta = \xi = 10^{-9}$ SI units; effective distance $d$=10 nm, $\varepsilon_{33}^b \le 5$.

Under the *absence of pinning*, thermodynamically equilibrium domain wall boundary $\rho(z)$ can be determined from the Eq.(5) under the condition $P_3(\rho, z) = 0$. In particular, the domain radius bias dependence $r(V)$ the sample surface should be determined from the equation $P_3(r, 0) = 0$, while the domain length $l(V)$ is determined from the equation $P_3(0, l) = 0$. These equations have nonzero roots under the condition $p_V(V)/P_S > 1$, namely at $L_\perp << d$ we derived:

$$r(V) \approx d\sqrt{\left(\frac{p_V(V)}{P_S}\right)^{2/3} - 1}, \qquad l(V) \approx \gamma_V(V)\, d\left(\left(\frac{p_V(V)}{P_S}\right)^{1/2} - 1\right). \tag{7}$$

At high $p_V(V)/P_S >> 1$, Eq.(7) leads to the relation $r^3/l^2 \approx \gamma_V^{-2}$ (compare with invariant $r^3/l^2 \approx \text{const}$ obtained within rigid approach by Molotskii [15]). So, the approximate expression for the critical bias $V_c$ of domain formation determined from the condition $p_V(V_c) = P_S$ is

$$V_c \approx d(\gamma + \gamma_b)\left(\frac{1}{2\varepsilon_0\varepsilon_{11}(\gamma + \gamma_b)^2} + \frac{1}{4d^2}\left(\eta + \frac{\xi}{\gamma^2}\right) + \frac{\alpha}{2} + 1.25\beta P_S^2 + 1.85\delta P_S^4\right). \tag{8}$$

Here $\gamma \equiv \gamma_V(0) \approx \sqrt{\varepsilon_{33}/\varepsilon_{11}}$ is dielectric anisotropy factor. The first term in brackets is depolarization energy contribution, the second term is domain wall energy, and last ones are nonlinear correlation energy contribution.

Let us underline, that domain radius $r(V) \sim d$ and length $l(V) \sim \gamma_V d$, calculated from Eqs.(7) are always finite, even in particular case of negligibly small domain wall energy (ξ=η=0) and so infinitely-thin domain walls ($L_\perp = L_z = 0$). This reflects the fact that spontaneous polarization re-orientation takes place inside the localized spatial region, where the resulting electric field absolute value is more that thermodynamic coercive field, i.e. $|E_3| > E_C$, while the hysteresis phenomenon appeared in the range $|E_3| < E_C$ as anticipated within LGD approach. Note, that nonlinear correlation energy $G_N$ contribution could dominate for some ferroelectrics and so effectively precludes rapid domain elongation with applied voltage increase.

Domain length $l(V)$ and radius $r(V)$ bias dependence calculated from Eqs.(6)-(7) within LGD-approach are shown in Figs.3 for typical ferroelectric materials. As anticipated, the aspect ratio of domain at initial growth stages ($V \geq V_c$) is close to the dielectric anisotropy factor γ. Calculated critical biases (2-20 V) of domain reversal are in reasonable agreement with available experimental results [3-8].

Note, that domain length decrease with applied bias increase (see Fig.3a at $V >> V_c$) may qualitatively explain anomalous polarization back-switching phenomena observed in LTO and PZT during the domain formation [30, 31, 32, 33]. Surely, for the adequate quantitative theoretical description of the back-switching pinning and strain effects should be considered.

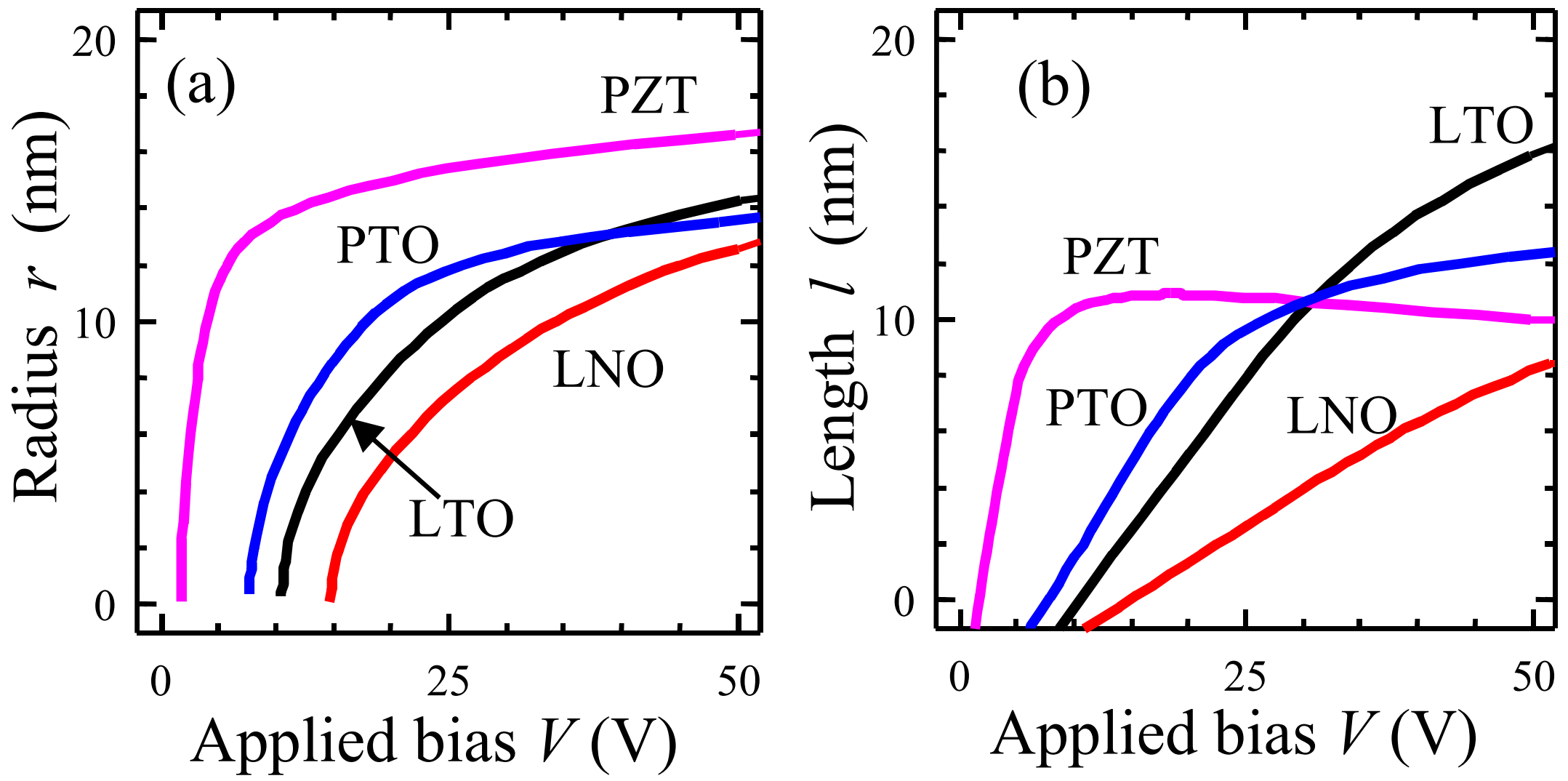

**FIG. 3**. Thermodynamically stable domain radius $r(V)$ (a) and length $l(V)$ (d) bias dependence calculated from Eqs.(6)-(7) within LGD- LGD approach combined with 2 parametric direct variational method for typical ferroelectric materials: PTO, PZT, LTO and LNO. Material parameters and probe characteristics are the same as in Fig.2.

The profiles of the probe-induced domain nucleus calculated within LGD approach from Eq.(5)-(6) are shown in Figs. 4. Under the bias increase polarization value increases in the region $\{r < d, z < d\}$ and domain formation starts at $V > V_c$.

Within LGD approach, oblate domain shape appears energetically preferable at $\gamma\sim 1$, since the domain wall energy and nonlinear correlation energy (both proportional to domain length $l\sim\gamma_V d$ in accordance with Eq.(6)) are relatively low, while their depolarization energy strongly decreases (up to $10^2$-$10^3$ times) with the intrinsic domain wall width increase from 0 to 1-2 nm.

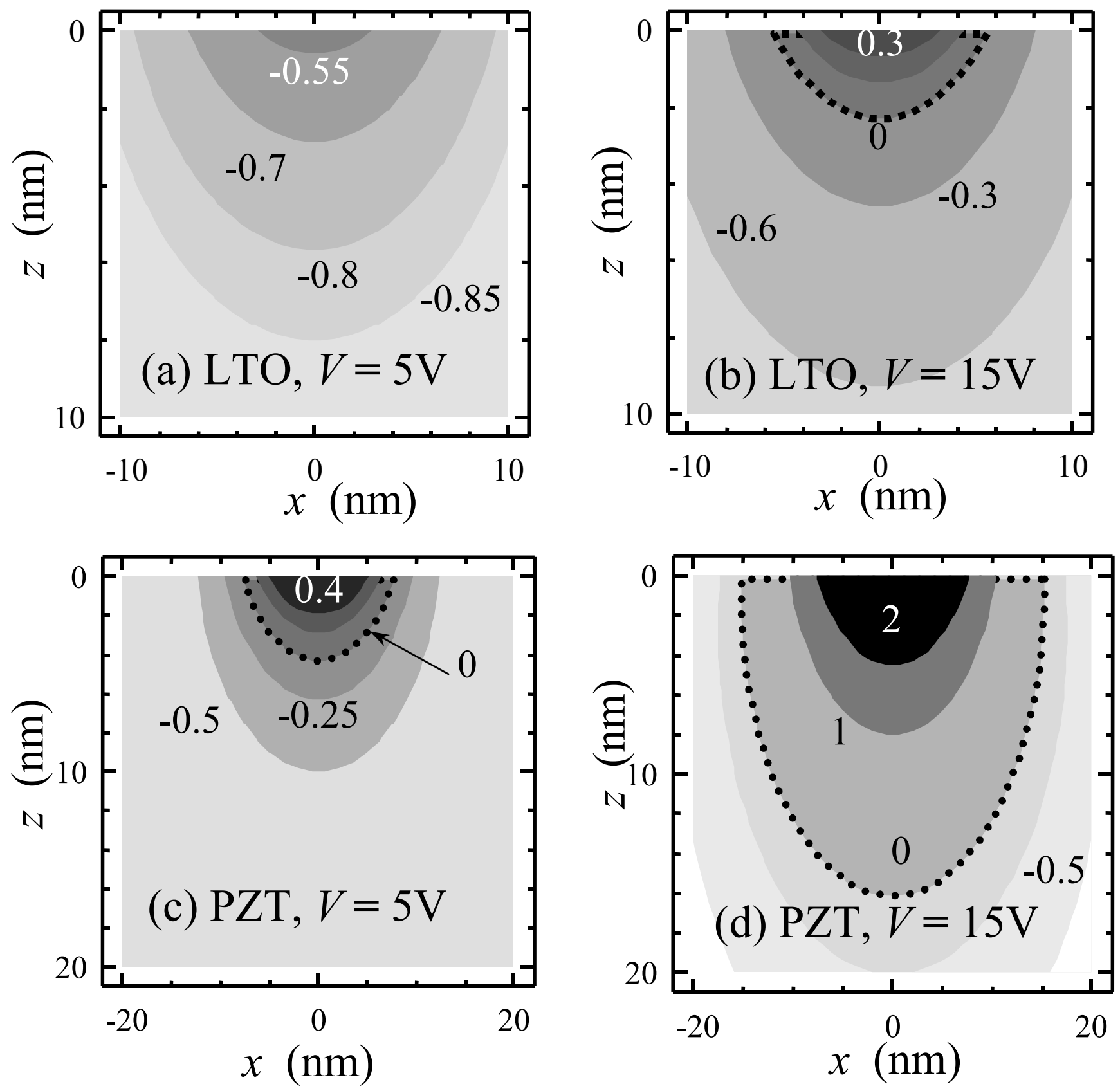

**FIG. 4**. Polarization distribution for LTO (a,b) and PZT (c,d) at applied bias 5 (a,c) and 15 V (b,d). Figures near the contours are polarization values in $P_S$ units. Dotted contour is the domain wall boundary $P_3(x,z)=0$. Material parameters and probe characteristics are the same as in Fig.2.

**5. Summary**

- We combine ***direct variational method with 2-parametric trial function*** with ***Landau-Ginzburg-Devonshire approach***, which considers intrinsic domain wall width and nonlinear correlation effects, for description of artificial ferroelectric nanodomain formation by using Scanning Probe Microscopy.

- We demonstrate that the shape of nanodomain induced by the probe can be either oblate or prolate depending on the nonlinearity strength. Calculated critical biases of domain reversal are in reasonable agreement with available experimental results. Spike-like domains (typical within Landauer-Molotskii rigid approach) are possible only when depolarization field energy contribution strongly dominates over the nonlinear field effects and domain wall energy.
- Using direct variational method with 2-parametric trial function, we derived analytical expressions are valid for both first and second order ferroelectrics. The expressions provide insight how the polarization re-distribution depends on the gradient energy, nonlinear correlation and depolarization effects, probe electrostatic potential distribution and ferroelectric material parameters.
- Note that high PFM response contrast is possible when reversed polarization value near the probe apex is several times higher than the sample spontaneous polarization far from the probe. This opens pathway for high-density data storage in ferroelectric materials with high nonlinear and correlation effects.

**Acknowledgements**

Author gratefully acknowledges Dr.Dr. E.A. Eliseev and Dr. Prof. S.V. Kalinin for fruitful discussions and valuable remarks, financial support from National Academy of Science of Ukraine, Ministry of Science and Education of Ukraine (UU30/004) and National Science Foundation (DMR-0908718 and DMR-0820404).

**Appendix**

Polarization distribution $P_3(\mathbf{r}) = -P_S + p(\mathbf{r})$ given by Eq.(5) should be substituted into the free energy:

$$G(P_3)=\frac{1}{Sh}\int_{-\infty}^{\infty}dx\int_{-\infty}^{\infty}dy\int_{0}^{h}dz\begin{pmatrix}\frac{\alpha}{2}P_3^2+\frac{\beta}{4}P_3^4+\frac{\delta}{6}P_3^6+\frac{\xi}{2}\left(\frac{\partial P_3}{\partial z}\right)^2\\+\frac{\eta}{2}(\nabla_\perp P_3)^2-P_3\left(E_3^e+\frac{E_3^d}{2}\right)\end{pmatrix}. \quad \text{(A.1)}$$

$S$ is the sample cross-section, external field $E_3^e(x,y,z)\sim V$. Electrical boundary conditions are $\int_0^h dzE_3=V_e(x,y)$ and $\int_0^h dzE_3^d=0$. Corresponding free energy excess is

$$\Delta G=\frac{2\pi}{S}\int_0^\infty\rho d\rho\int_0^h dz\begin{pmatrix}\frac{\alpha}{2}\left(p^2-2P_S p\right)+\frac{\beta}{4}\left(p^4-4P_S p^3+6P_S^2p^2-4P_S^3p\right)+\\ \frac{\delta}{6}\left(p^6-6P_S p^5+15P_S^2p^4-20P_S^3p^3+15P_S^4p^2-6P_S^5p\right)\\ +\frac{\xi}{2}\left(\frac{\partial p}{\partial z}\right)^2+\frac{\eta}{2}\left(\frac{\partial p}{\partial\rho}\right)^2-p\left(E_3^e+\frac{1}{2}E_3^d[p]\right)\end{pmatrix}. \quad \text{(A.2)}$$

At $h\to\infty$ the external field is $E_3^e(\rho,z)\approx\frac{V(d+z/\gamma_b)d}{\gamma_b\left((d+z/\gamma_b)^2+\rho^2\right)^{3/2}}$, where $\gamma_b=\sqrt{\varepsilon_{33}^b/\varepsilon_{11}}$. Since $P_S^2=-\alpha/\beta$ for the second order ferroelectric, while $P_S^2=\left(\sqrt{\beta^2-4\alpha\delta}-\beta\right)/2\delta$ for the first order one, we obtains:

$$\begin{pmatrix}\frac{\alpha}{2}\left(p^2-2P_S p\right)+\frac{\beta}{4}\left(p^4-4P_S p^3+6P_S^2p^2-4P_S^3p\right)+\\ \frac{\delta}{6}\left(p^6-6P_S p^5+15P_S^2p^4-20P_S^3p^3+15P_S^4p^2-6P_S^5p\right)\end{pmatrix}=\begin{pmatrix}\frac{\alpha_P}{2}p^2-\left(\beta+\frac{10\delta}{3}P_S^2\right)P_S p^3+\\ \left(\frac{\beta}{4}+\frac{5\delta}{2}P_S^2\right)p^4-\delta P_S p^5+\frac{\delta}{6}p^6\end{pmatrix}.$$

Notice, that coefficient $\alpha_P=\alpha+3\beta P_S^2+5\delta P_S^4$ is always positive.

The free energy expression for the second order ferroelectrics was derived as

$$G(p_V,\gamma_V)=2\pi\left(Q_1(\gamma_V,V)p_V+Q_2(\gamma_V)p_V{}^2-Q_3(\gamma_V)p_V{}^3+Q_4(\gamma_V)\frac{p_V{}^4}{4}\right). \quad \text{(A.3)}$$

Where:

$$Q_1(\gamma_V,V)\approx -Vd(L_\perp+d)\frac{\gamma_V}{2(\gamma_V+\gamma_b)}\quad \text{for}\quad L_\perp<<d, \quad \text{(A.4a)}$$

$$Q_2(\gamma_V)=\begin{pmatrix} -\alpha\frac{\gamma_V d^2}{2L_\perp^2}(L_\perp+d)^2\left(L_\perp-d\cdot\ln\left(1+\frac{L_\perp}{d}\right)\right)+\gamma_V\frac{\eta d}{16}+\frac{\xi d}{16\gamma_V}\left(1+\frac{L_\perp}{6d}\right) \\ +\frac{d}{8}\frac{(L_\perp+d)^2}{\varepsilon_0\varepsilon_{33}^b}\frac{\gamma_V}{(\gamma_V+\gamma_b)^2} \end{pmatrix} \qquad \text{(A.4b)}$$

$$\approx -\alpha\frac{\gamma_V d^3}{4}\left(1+\frac{4L_\perp}{3d}\right)+\frac{d^3}{8}\frac{\gamma_V(1+2L_\perp/d)}{\varepsilon_0\varepsilon_{33}^b(\gamma_V+\gamma_b)^2}+\frac{\gamma_V d}{16}\left(\eta+\frac{\xi}{\gamma_V^2}\left(1+\frac{L_\perp}{6d}\right)\right) \quad \text{for} \quad L_\perp \ll d$$

$$Q_3(\gamma_V)=\beta P_S\frac{\gamma_V d^3}{4L_\perp^3}(L_\perp+d)^3\begin{pmatrix} 4\ln\left(1+\frac{L_\perp}{d}\right)+\frac{L_\perp}{L_\perp+d}+ \\ 15\left(\sqrt{\frac{d}{L_\perp}}\arccos\left(\sqrt{\frac{d}{L_\perp+d}}\right)-1\right) \end{pmatrix}, \qquad \text{(A.4c)}$$

$$\approx \beta P_S\frac{\gamma_V d^3}{21}\left(1+\frac{5L_\perp}{4d}\right) \quad \text{for} \quad L_\perp \ll d$$

$$Q_4(\gamma_V)=\beta\frac{\gamma_V d^3}{12L_\perp^5}(L_\perp+d)^2\begin{pmatrix} d(L_\perp+d)^2\left(26\ln\left(1+\frac{L_\perp}{d}\right)+24\,\mathrm{Li}_2\left(-\frac{L_\perp}{d}\right)\right) \\ -L_\perp\left(2d^2-3L_\perp d-6L_\perp^2\right) \end{pmatrix}, \qquad \text{(A.4d)}$$

$$\approx \beta\frac{\gamma_V d^3}{50}\left(1+\frac{11L_\perp}{9d}\right) \quad \text{for} \quad L_\perp \ll d$$

Where the polylogarithm function $\mathrm{Li}_n(x)=\sum_{k=1}^{\infty}x^k/k^n$ . Expression (6) listed in the main text for the first order ferroelectrics was obtained by the similar way, but much more cumbersome.

In the limiting case $L_\perp \ll d$ , one obtains the free energy expression

$$G(p_V,\gamma_V)\approx 2\pi\begin{pmatrix} \left(\frac{\gamma_V d^3}{2\varepsilon_0\varepsilon_{11}(\gamma_V+\gamma_b)^2}+\frac{\gamma_V d}{4}\left(\eta+\frac{\xi}{\gamma_V^2}\right)-\alpha\right)\frac{p_V{}^2}{4} \\ -\beta P_S\gamma_V d^3\frac{p_V{}^3}{21}+\frac{\beta}{50}\gamma_V d^3\frac{p_V{}^4}{4}-\frac{Vd^2\gamma_V}{2(\gamma_V+\gamma_b)}p_V \end{pmatrix}. \qquad \text{(A.5)}$$

Equations of state $\partial G/\partial\gamma_V=0$, $\partial G/\partial p_V=0$ are:

$$\left(\frac{1}{2\varepsilon_0\varepsilon_{11}(\gamma_V+\gamma_b)^2}+\frac{1}{4d^2}\left(\eta+\frac{\xi}{\gamma_V^2}\right)-\alpha\right)p_V-\frac{2}{7}\beta P_S p_V{}^2+\frac{\beta}{25}p_V{}^3=\frac{V}{d(\gamma_V+\gamma_b)}, \qquad \text{(A.6a)}$$

At $p_V \neq 0$

$$\left(\frac{\eta}{4}-\frac{\xi}{4\gamma_V^2}-\frac{d^2(\gamma_V-\gamma_b)}{2\varepsilon_0\varepsilon_{11}(\gamma_V+\gamma_b)^3}-\alpha d^2\right)p_V-\frac{4d^2}{21}\beta P_S\,p_V^2+\frac{d^2}{50}\beta p_V^3=\frac{2Vd\gamma_b}{(\gamma_V+\gamma_b)^2}. \quad \text{(A.6b)}$$

Under the condition $\gamma_b << \gamma_V$, Eq.(A.6b) can be easily solved. Then he domain radius bias dependence $r(V)$ the sample surface should be determined from the equation as $P_3(r,0)=0$, while the domain length $l(V)$ is determined from the equation $P_3(0,l)=0$. Namely we derived:

$$\begin{cases}\left(L_\perp d+d^2+r^2\right)\sqrt{d^2+r^2}=d^2\left(L_\perp+d\right)\dfrac{p_V}{P_S},\\ L_\perp\left(d+l/\gamma_V\right)+\left(d+l/\gamma_V\right)^2=d\left(L_\perp+d\right)\dfrac{p_V}{P_S}.\end{cases} \quad \text{(A.7)}$$

Under the condition $p_V/P_S>1$ and $L_\perp << d$, the equations (A.7) have nonzero roots given by Eq.(7) in the main text. The approximate expression for the critical bias $V_c$ determined from the condition $p_V(V_c)=P_S$ is

$$V_c=2d(\gamma_V+\gamma_b)\left(\begin{array}{l}\left(\dfrac{1}{4\varepsilon_0\varepsilon_{11}(\gamma_V+\gamma_b)^2}+\dfrac{1}{8d^2}\left(\eta+\dfrac{\xi}{\gamma_V^2}\right)+\dfrac{\alpha+3\beta P_S^2+5\delta P_S^4}{4}\right)\\ -\left(\dfrac{\beta}{7}P_S^2+\dfrac{10\delta}{21}P_S^4\right)+\left(\dfrac{\beta}{50}P_S^2+\dfrac{\delta}{5}P_S^4\right)-\dfrac{5\delta}{91}P_S^4+\dfrac{\delta}{144}P_S^4\end{array}\right). \quad \text{(A.8)}$$

## References

---